\documentclass[american,english]{revtex4}
\usepackage[T1]{fontenc}
\usepackage[latin9]{inputenc}
\usepackage{color}
\usepackage{float}
\usepackage{textcomp}
\usepackage{multirow}
\usepackage{amsmath}
\usepackage{amssymb}
\usepackage{graphicx}

\makeatletter

\newcommand{\lyxmathsym}[1]{\ifmmode\begingroup\def\b@ld{bold}
  \text{\ifx\math@version\b@ld\bfseries\fi#1}\endgroup\else#1\fi}

\providecommand{\tabularnewline}{\\}

\@ifundefined{textcolor}{}
{%
 \definecolor{BLACK}{gray}{0}
 \definecolor{WHITE}{gray}{1}
 \definecolor{RED}{rgb}{1,0,0}
 \definecolor{GREEN}{rgb}{0,1,0}
 \definecolor{BLUE}{rgb}{0,0,1}
 \definecolor{CYAN}{cmyk}{1,0,0,0}
 \definecolor{MAGENTA}{cmyk}{0,1,0,0}
 \definecolor{YELLOW}{cmyk}{0,0,1,0}
}

\makeatother

\usepackage{babel}
\begin{document}
\title{Global Natural Orbital Functional:\\
Towards the Complete Description of the Electron Correlation}
\author{Mario Piris }
\address{Kimika Fakultatea, Euskal Herriko Unibertsitatea (UPV/EHU), P.K. 1072, 20080 Donostia, Spain.}
\address{Donostia International Physics Center (DIPC), 20018 Donostia, Spain.}
\address{IKERBASQUE, Basque Foundation for Science, 48013 Bilbao, Spain.}

\begin{abstract}
The current work presents a natural orbital functional (NOF) for electronic systems with any spin value independent of the external potential being considered, that is, a global NOF (GNOF). It is based on a new two-index reconstruction of the two-particle reduced density matrix for spin multiplets. The emergent functional describes the complete intrapair electron correlation, and the correlation between orbitals that make up both the pairs and the individual electrons. The interorbital correlation is composed of static and dynamic terms. The concept of dynamic part of the occupation numbers is introduced. To evaluate the accuracy achieved with GNOF, calculation of a variety of properties is presented. They include the total energies and energy differences between the ground state and the lowest-lying excited state with different spin of atoms from H to Ne, ionization potentials of the first-row transition-metal atoms (Sc-Zn), and the total energies of a selected set of 55 molecular systems in different spin states. GNOF is also applied to the homolytic dissociation of selected diatomic molecules in different spin states and to the rotation barrier of ethylene, both paradigmatic cases of systems with significant multi-configurational character. The values obtained agree with those reported at high level of theory and experimental data.
\end{abstract}

\maketitle

In the non-relativistic limit, for time-independent and spin-free Hamiltonians, the ground state of a many-electron system with spin $S$ is a multiplet, that is, a mixture of pure states with all possible spin projections. Such an ensemble is represented by its associated density matrix $\mathfrak{D^{\mathrm{N}}}$ which allows describing quantum observables through statistical averages. Unfortunately, the number of variables involved in determining $\mathfrak{D^{\mathrm{N}}}$ grows astronomically with the number N of electrons, and actually contains significantly more information than is necessary to calculate energies and properties.

Appropriate representations of the electronic structure of atoms, molecules and solids without explicit recourse to $\mathfrak{D^{\mathrm{N}}}$ can alternatively be obtained by the one-particle reduced density matrix (1RDM) functional theory \citep{Gilbert1975}. Here, the 1RDM $\Gamma$, a much simpler object than $\mathfrak{D^{\mathrm{N}}}$, is used directly for ground-state variational calculations. Valone proved \citep{Valone1980} the existence of the functional $E\left[\Gamma\right]$ for ensembles by extending Levy's functional \citep{Levy1979} to all ensemble N-representable 1RDMs \citep{Coleman1963}. $E\left[\Gamma\right]$ is defined independently of the external potential under consideration and is therefore a universal functional. Regrettably, computational schemes based on the exact constrained search formulation are too expensive; so the 1RDM functional requires a practical approach.

It is well known that the ground-state energy of an N-particle quantum system with a Hamiltonian involving not more than two-body interactions can be cast as an exact functional $E\left[\mathrm{D}\right]$
of the two-particle reduced density matrix (2RDM) D. Hence, the 1RDM functional $E\left[\Gamma\right]$ must match the 2RDM functional $E\left[\mathrm{D}\right]$. Actually, we must only reconstruct the
electron-electron potential energy $V_{ee}$ in terms of the 1RDM since the non-interacting part of the electronic Hamiltonian is a one-particle operator. Unfortunately, the explicit reconstruction
$V_{ee}\left[\Gamma\right]$ has resulted in unattainable goal so far, and we have to settle for making approximations.

The typical approach is to employ the exact $V_{ee}\left[\mathrm{D}\right]$ but using solely a reconstruction functional $\mathrm{D}\left[\Gamma\right]$. In general, the exact ground-state energy is not completely reconstructed, and approximate 2RDMs lead to functionals that are still implicitly dependent on D. An unwanted implication of this 2RDM dependence is that the functional N-representability problem arises \citep{Ludena2013,Piris2018}, i.e., a reconstructed D must be ensemble N-representable \citep{Mazziotti2012,Mazziotti2012b} as well. Otherwise, the approximate functional $V_{ee}\left[\Gamma\right]$ can lead to non-physical energy values.

The functionals currently in use are constructed in the basis where the 1RDM is diagonal which is the definition of a natural orbital functional (NOF). Hence, the electronic energy is expressed in terms of the natural orbitals (NOs) and their occupation numbers (ONs) $\left\{ n_{i}\right\} $, 

\begin{equation} 
E=\sum\limits _{i}n_{i}H_{ii}+\sum\limits _{ijkl}D[n_{i},n_{j},n_{k},n_{l}]<kl|ij>\label{ENOF}
\end{equation}

In Eq.$\,$(\ref{ENOF}), $H_{ii}$ denotes the diagonal one-electron matrix elements of the kinetic energy and external potential operators, $<kl|ij>$ are the matrix elements of the two-particle interaction, and $D[n_{i},n_{j},n_{k},n_{l}]$ represents the reconstructed 2RDM from the ONs. Restriction on the ONs to the range $0\leq n_{i}\leq1$ represents a necessary and sufficient condition for ensemble N-representability of the 1RDM under the normalization condition $\sum_{i}n_{i}=\mathrm{N}$ \citep{Coleman1963}. A detailed account of the state of the art of the NOF theory (NOFT) can be found elsewhere \citep{Piris2007,Piris2014a,Pernal2016,Schade2017,Mitxelena2019}.

Several approximate functionals have been proposed \citep{Muller1984,Goedecker1998,Csanyi2000,Gritsenko2005,Sharma2008,Marques2008,Rohr2008}, but none of them guarantee that physical conditions such as 2RDM antisymmetry are preserved \citep{Rodriguez-Mayorga2017}. Solely PNOFs \citep{Piris2013b,Piris2014c,Piris2017}, which are based on the reconstruction of $D$ subject to necessary N-representability conditions, can guarantee this. These functionals are capable of producing a qualitatively correct description of systems with a multiconfigurational nature, one of the greatest challenges for density functionals, achieving chemical accuracy in many cases \citep{Mitxelena2020,Mitxelena2020b}. Nevertheless, they also suffer from an important lack of dynamic correlation. To recover this correlation,
second-order perturbative corrections have been implemented with significant results \citep{Piris2013c,Piris2018b,Lopez2019,Mercero2021}. In this work, however, it is intended to recover the missing dynamic correlation within the NOFT framework only.

The goal is to design an accurate NOF for all electronic structure problems, that is, a global NOF (GNOF). We limit ourselves to two-index reconstruction $D[n_{i},n_{j}]$, aimed at obtaining
the least possible scaling with the size of the system. It is worth noting that the adjective `global' is used instead of `universal' to differentiate our multipurpose approximate NOF from the Valone's
exact.

Consider that $\mathrm{N_{I}}$ single electrons determine the spin of the system, $S=\mathrm{N_{I}}/2$, and the rest of electrons, $\mathrm{N_{II}}=\mathrm{N-N_{I}}$, are spin-paired providing zero spin. We focus on the mixed state of highest multiplicity: $2S+1=\mathrm{N_{I}}+1$. Then, $\mathrm{<}\hat{S}_{z}\mathrm{>=0}$ for the whole ensemble \citep{Piris2019}, so we can adopt the spin-restricted
formalism in which a single set of orbitals is used for $\alpha$ and $\beta$ spins. All spatial orbitals will be then double occupied, so that ONs for particles with $\alpha$ and $\beta$ spins are equal:
$n_{p}^{\alpha}=n_{p}^{\beta}=n_{p}.$

Next, divide the orbital space $\Omega$ into two subspaces: $\Omega=\Omega_{\mathrm{I}}\oplus\Omega_{\mathrm{II}}$. Both $\Omega_{\mathrm{I}}$ and $\Omega_{\mathrm{II}}$ are composed of $\mathrm{N_{I}}$ and $\mathrm{N_{II}}/2$ mutually disjoint subspaces $\Omega{}_{g}$, respectively. Each subspace $\Omega{}_{g}\in\Omega_{\mathrm{I}}$ contains only one orbital $g$ with $n_{g}=1/2$ which is individually
occupied, but we do not know whether the electron has $\alpha$ or $\beta$ spin. In contrast, each $\Omega{}_{g}\in\Omega_{\mathrm{II}}$ is double occupied and contains one orbital with $g\leq\mathrm{N_{II}}/2$, and $\mathrm{N}_{g}$ orbitals \{$\phi_{p}$\}=\{$\phi_{p_{1}},\phi_{p_{2}},...,\phi_{pN_{g}}$\} with $p>\mathrm{N}_{\Omega}=\mathrm{N_{II}}/2+\mathrm{N_{I}}$. Find
an illustrative example of splitting into subspaces in Fig. 1 of the supplementary material (SM). Taking into account the spin, the trace of the 1RDM is verified \citep{Piris2019} equal to the number of
electrons ($2\sum n_{p}=\mathrm{N}$).

Now it is time to rebuild the 2RDM from the ONs. We divide D into intra- and inter-subspace contributions. For intra-subspace blocks, we consider only intrapair contributions:

\begin{equation}
\begin{array}{c}
D_{pq,rt}^{\alpha\beta}=\left[{\displaystyle \frac{n_{p}\delta_{pr}+\Pi_{pr}\left(1-\delta_{pr}\right)}{2}}\right]\delta_{pq}\delta_{rt}\delta_{p\Omega_{g}}\delta_{r\Omega_{g}}\quad\\
\Pi_{pr}=-\sqrt{n_{p}n_{r}}\left(\delta_{pg}+\delta_{rg}-\delta_{p\Omega^{a}}\delta_{r\Omega^{a}}\right),\;g\leq\mathrm{N_{II}}/2
\end{array}\label{intra}
\end{equation}

Note that $D_{pp,pp}^{\alpha\beta}=0$ if $p\in\Omega_{\mathrm{I}}$ since there can be only one electron with $\alpha$ or $\beta$ spin in each pure state $\left|SM_{s}\right\rangle $ of the ensemble \citep{Piris2019}. Kronecker deltas have an obvious meaning, for instance, $\delta_{p\Omega_{g}}=1$ if $p\in\Omega_{g}$ or $\delta_{p\Omega_{g}}=0$ otherwise. $\Omega^{a}=\Omega_{\mathrm{II}}^{a}$
denotes the subspace composed of orbitals above the level $\mathrm{N}_{\Omega}$ ($p>\mathrm{N}_{\Omega}$). Reconstruction ($\ref{intra})$ in Eq. (\ref{ENOF}) leads to PNOF5 \citep{Piris2013e}, a sum of $\mathrm{N_{II}}$/2 pair energies accurately described by the Löwdin's venerable two-electron functional.

For inter-subspace contributions ($\Omega{}_{f}\neq\Omega{}_{g}$), the spin-parallel blocks are assummed Hartree-Fock (HF) like,

\begin{equation}
D_{pq,rt}^{\sigma\sigma}={\displaystyle \frac{n_{p}n_{q}}{2}}\left(\delta_{pr}\delta_{qt}-\delta_{pt}\delta_{qr}\right)\delta_{p\Omega_{f}}\delta_{q\Omega_{g}},\:{}_{\sigma=\alpha,\beta}\label{interaa}
\end{equation}

whereas the spin-antiparallel blocks are taken as

\begin{equation}
\begin{array}{c}
D_{pq,rt}^{\alpha\beta}=\left[{\displaystyle {\displaystyle \frac{n_{p}n_{q}}{2}\delta_{pr}\delta_{qt}-\frac{\delta_{p\Omega_{\mathrm{I}}}\delta_{q\Omega_{\mathrm{I}}}}{8}\delta_{pt}\delta_{qr}}}\right]\delta_{p\Omega_{f}}\delta_{q\Omega_{g}}
-{\displaystyle \frac{{\displaystyle \Pi_{pr}^{s}}+{\displaystyle \Pi_{pr}^{d}}}{2}}\delta_{pq}\delta_{rt}\delta_{p\Omega_{f}}\delta_{r\Omega_{g}}\qquad\\
\\
\Pi_{pr}^{s}=\sqrt{n_{p}(1-n_{p})n_{r}(1-n_{r})}\:[\:\delta_{p\Omega^{b}}\delta_{r\Omega^{a}}+\delta_{p\Omega^{a}}\delta_{r\Omega^{b}}+\delta_{p\Omega^{a}}\delta_{r\Omega^{a}}+\tfrac{1}{2}(\delta_{p\Omega_{\mathrm{II}}^{b}}\delta_{r\Omega_{\mathrm{I}}}+\delta_{p\Omega_{\mathrm{I}}}\delta_{r\Omega_{\mathrm{II}}^{b}})]\\
\\
\Pi_{pr}^{d}=\left(\sqrt{n_{p}^{d}n_{r}^{d}}-n_{p}^{d}n_{r}^{d}\right)\left(\delta_{p\Omega_{\mathrm{II}}^{b}}\delta_{r\Omega^{a}}+\delta_{p\Omega^{a}}\delta_{r\Omega_{\mathrm{II}}^{b}}\right)
-\left(\sqrt{n_{p}^{d}n_{r}^{d}}+n_{p}^{d}n_{r}^{d}\right)\delta_{p\Omega^{a}}\delta_{r\Omega^{a}}\qquad\qquad\;
\end{array}\label{interab}
\end{equation}

where $\Omega^{b}\equiv p\leq\mathrm{N}_{\Omega}$ and $\Omega_{\mathrm{II}}^{b}\equiv p\leq\mathrm{N_{II}}/2$. Observe that interactions between orbitals belonging to $\Omega_{\mathrm{II}}^{b}$
are not considered in $\Pi$ matrices. It is worth noting that Eqs. ($\ref{intra})$ - ($\ref{interab})$ satisfy some analytical conditions necessary for the ensemble N-representability of the 2RDM, as in the preceding PNOFs.

$\Pi^{s}$ and $\Pi^{d}$ are responsible for the static and dynamic correlation between subspaces, respectively, in accordance with the Pulay's criterion that establishes an occupancy deviation of approximately 0.01 with respect to 1 or 0 for a NO to contribute to the dynamic correlation, while larger deviations contribute to the non-dynamic correlation. For $\Pi^{s}$, the PNOF7 functional
form \citep{Piris2017} has been adopted, hence its square root has significant values only when the ONs differ substantially from 1 and 0.

\begin{figure}
\centering{}\includegraphics[scale=0.3]{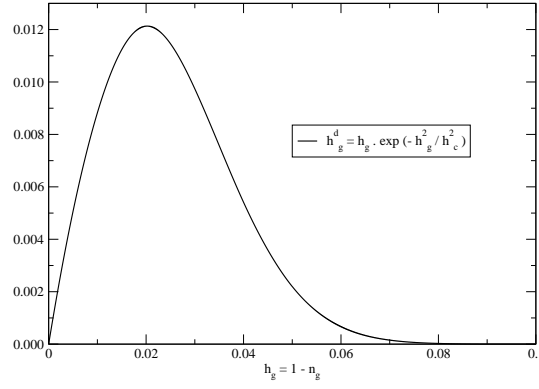}\caption{\label{fig2} Dynamic hole $h_{g}^{d}$ for $h_{c}=0.02\sqrt{2}$.
}
\end{figure}

Taking into account that $\Pi$ in Eq. (\ref{intra}) is capable of recovering the whole intrapair correlation, the functional form of $\Pi^{d}$ is expected to be proportional to the product of the square roots of the ONs when these correspond to very small deviations. Let us define the dynamic part of $n_{p}$ as
\begin{equation}
n_{p}^{d}=n_{p}\cdot\dfrac{h_{g}^{d}}{h_{g}}\;,\quad p\in\Omega_{g}\,,\quad g=1,2,...,{\displaystyle \mathrm{N_{II}}/2}\label{dyn-on}
\end{equation}
The hole $h_{g}=1-n_{g}$, while its dynamic part reads as
\begin{equation}
h_{g}^{d}=h_{g}\cdot e^{-\left(\dfrac{h_{g}}{h_{c}}\right)^{2}},\quad g=1,2,...,{\displaystyle \mathrm{N_{II}}/2}\label{dyn-hole1-1}
\end{equation}
In Fig. \ref{fig2}, $h_{g}^{d}$ is shown for $h_{c}=0.02\sqrt{2}$.
The maximum value is around 0.012, in accordance with the Pulay's
criterion. Considering real spatial orbitals and $n_{p}\approx n_{p}^{d}$,
it is not difficult to verify that the terms proportional to the product
of the ONs in $D^{\alpha\beta}$ will cancel out with the corresponding
terms of $D^{\sigma\sigma}$ in the energy expression (\ref{ENOF}),
so that only those terms proportional to $\sqrt{n_{p}^{d}n_{r}^{d}}$
will contribute to the energy.

Substituting in Eq. (\ref{ENOF}) the expressions (\ref{intra}), (\ref{interaa}) and (\ref{interab}) for the 2RDM blocks, the GNOF is obtained:

\begin{equation}
E=E^{intra}+E_{HF}^{inter}+E_{sta}^{inter}+E_{dyn}^{inter}
\end{equation}
\begin{equation}
\begin{array}{c}
E^{intra}=\sum\limits _{g=1}^{\mathrm{N_{II}}/2}E_{g}+{\displaystyle \sum_{g=\mathrm{N_{II}}/2+1}^{\mathrm{N}_{\Omega}}}H_{gg}\quad,\qquad
E_{g}=\sum\limits _{p\in\Omega_{g}}n_{p}(2H_{pp}+J_{pp}) + \sum\limits _{q,p\in\Omega_{g},p\neq q}\Pi_{qp}L_{pq}
\end{array}
\end{equation}
\begin{equation}
E_{HF}^{inter}=\sum\limits _{p,q=1}^{\mathrm{N}_{B}}\,'\,n_{q}n_{p}\left(2J_{pq}-K_{pq}\right)
\end{equation}
\begin{equation}
\begin{array}{c}
E_{sta}^{inter}=-\left({\displaystyle \sum_{p=1}^{\mathrm{N}_{\Omega}}\sum_{q=\mathrm{N}_{\Omega}+1}^{\mathrm{N}_{B}}+\sum_{p=\mathrm{N}_{\Omega}+1}^{\mathrm{N}_{B}}\sum_{q=1}^{\mathrm{N}_{\Omega}}}\right.
\left.{\displaystyle +\sum_{p,q=\mathrm{N}_{\Omega}+1}^{\mathrm{N}_{B}}}\right)'\sqrt{n_{q}h_{q}n_{p}h_{p}}L_{pq}\qquad\qquad\\
\qquad\qquad\quad-\:\dfrac{1}{2}\left({\displaystyle \sum\limits _{p=1}^{\mathrm{N_{II}}/2}\sum_{q=\mathrm{N_{II}}/2+1}^{\mathrm{N}_{\Omega}}+\sum_{p=\mathrm{N_{II}}/2+1}^{\mathrm{N}_{\Omega}}\sum\limits _{q=1}^{\mathrm{N_{II}}/2}}\right)' \sqrt{n_{q}h_{q}n_{p}h_{p}}L_{pq}{\displaystyle \:-\:\dfrac{1}{4}\sum_{p,q=\mathrm{N_{II}}/2+1}^{\mathrm{N}_{\Omega}}}K_{pq}
\end{array}
\end{equation}
\begin{equation}
\begin{array}{c}
E_{dyn}^{inter}=-\left(\sum\limits _{p=1}^{\mathrm{N_{II}}/2}\sum\limits _{q=\mathrm{N}_{\Omega}+1}^{\mathrm{N}_{B}}+\sum\limits _{p=\mathrm{N}_{\Omega}+1}^{\mathrm{N}_{B}}\sum\limits _{q=1}^{\mathrm{N_{II}}/2}\right)'\left({\displaystyle \sqrt{n_{q}^{d}n_{p}^{d}}}\right.
\left.-n_{q}^{d}n_{p}^{d}\right)L_{pq}+\sum\limits _{p,q=\mathrm{N}_{\Omega}+1}^{\mathrm{N}_{B}}\,'\left(\sqrt{n_{q}^{d}n_{p}^{d}}+n_{q}^{d}n_{p}^{d}\right)L_{pq}
\end{array}
\end{equation}

where $J_{pq}=\left\langle pq|pq\right\rangle $, $K_{pq}=\left\langle pq|qp\right\rangle $, and $L_{pq}=\left\langle pp|qq\right\rangle $ are the Coulomb, exchange, and exchange-time-inversion integrals \citep{Piris1999}, respectively. $\mathrm{N}_{B}$ denotes the number of basic functions considered. In the summations, the prime indicates that only the inter-subspace terms are taking into account ($p\in\Omega{}_{f},q\in\Omega{}_{g},f\neq g$). The simplified energy expression of GNOF in the case of singlet states can be found in the SM.

GNOF has the ability to retrieve the complete intrapair electron correlation and introduces interaction terms between orbitals that make up both the pairs and the individual electrons. The interorbital
correlation is in turn composed of the sum of the static and dynamic terms. It is not difficult to verify \citep{Piris2019} that $\mathrm{<}\hat{S}^{2}\mathrm{>}=S\left(S+1\right)$ as well. The solution is established by optimizing the energy with respect to the ONs and to the NOs, separately. Therefore, orbitals vary along the optimization process until the most favorable orbital interactions are found. All calculations have been carried out using the DoNOF code \citep{Piris2021} where the GNOF has been implemented. The procedure is simple, showing a formal scaling of $\mathrm{N}_{B}^{5}$ ($\mathrm{N}_{B}$: number of basis functions).

To measure the success of GNOF, calculation of a variety of properties is presented. The correlation-consistent valence triple-$\zeta$ basis set (cc-pVTZ) \citep{Pritchard2019} was used throughout, except in some cases that will be specified. For comparison, CCSD(T) values are reported obtained using the GAUSSIAN03 program package \citep{g03}. The experimental data come from the National Institute of Standards and Technology (NIST) Database \citep{NIST}. For experimental dissociation energies, it was also combined with the Ref. \citep{Chase1998}. It is not intended to reproduce the experimental
data in this work, since it requires large basis sets.

\begin{table}[H]
\centering{}\caption{\label{table1} Total energies (Hartrees) and the excitation energies (eV) of the lowest-lying excited state with different spin.}
\bigskip{}
\begin{tabular}{c|ccc|cccc}
\hline 
\multirow{1}{*}{At} & GS & GNOF & CCSD(T) & ES & GNOF & CCSD(T) & Exp\tabularnewline
\hline 
H   & $^{2}S$ &  -0.49983 &  -0.49983 &     -   &    -   &    -   &    -   \tabularnewline
He  & $^{1}S$ &  -2.90084 &  -2.90084 & $^{3}S$ &  19.91 &  19.88 &  19.82 \tabularnewline
Li  & $^{2}S$ &  -7.45318 &  -7.45338 & $^{4}P$ &  57.00 &  56.86 &  57.47 \tabularnewline
Be  & $^{1}S$ & -14.63382 & -14.63565 & $^{3}P$ & \ 2.76 & \ 2.72 & \ 2.72 \tabularnewline
B   & $^{2}P$ & -24.60751 & -24.60912 & $^{4}P$ & \ 3.83 & \ 3.55 & \ 3.58 \tabularnewline
C   & $^{3}P$ & -37.79635 & -37.79712 & $^{1}D$ & \ 1.52 & \ 1.43 & \ 1.26 \tabularnewline
N   & $^{4}S$ & -54.52947 & -54.53421 & $^{2}D$ & \ 2.20 & \ 2.72 & \ 2.38 \tabularnewline
O   & $^{3}P$ & -75.00049 & -74.99967 & $^{1}D$ & \ 2.28 & \ 2.21 & \ 1.97 \tabularnewline
F   & $^{2}P$ & -99.65391 & -99.65218 & $^{4}P$ &  13.33 &  13.34 &  12.70 \tabularnewline
Ne  & $^{1}S$ & -128.8442 & -128.8440 & $^{3}P$ &  17.70 &  17.78 &  16.62 \tabularnewline
\hline 
\footnotesize{}MAE &  & 0.0012 & - &  & \ 0.37 & \ 0.36 & - \tabularnewline
\hline 
\end{tabular}
\end{table}

Table \ref{table1} collects the total energies and energy differences between the ground state (GS) and the lowest-lying excited state (ES) with different spin for atoms from H to Ne. The aug-cc-pVTZ basis set was used \citep{Pritchard2019}. Experimental data is from Ref. \citep{NIST-ASD}. According to the mean absolute error (MAE), GNOF provides GS total energies w.r.t. CCSD(T) ones within the chemical accuracy (1 kcal/mol) for these atoms, whereas for excitation energies, both theoretical methods present MAEs w.r.t. the experiment that differ from each other by less than 1 kcal/mol too. Hence, GNOF provides these excitation energies with respect to experimental data comparably to CCSD(T). Recall that CCSD(T) employs an unrestricted formalism for non-singlet states, while GNOF preserves the total spin of the multiplet, therefore, excitation energies between states with different spin provided by both methods differ, namely for Li, N , F and Ne, GNOF provides values closer to the experiment, while CCSD (T) does better for He, Be, B, C and O.

\begin{figure}[t]
\centering{}\includegraphics[scale=0.4]{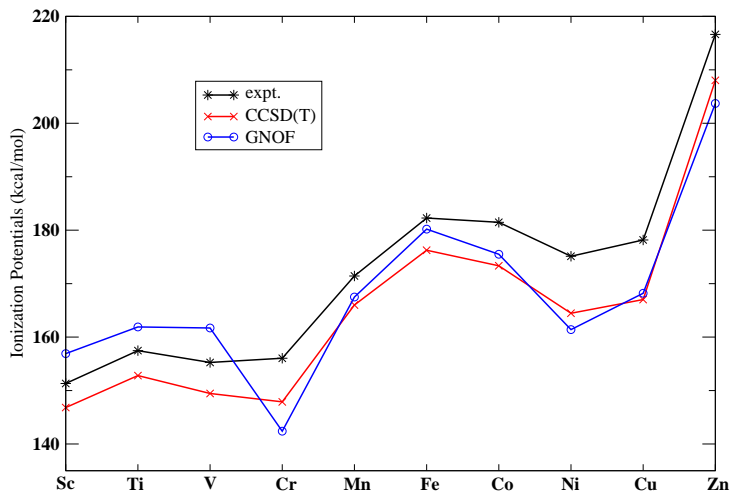}\caption{\label{fig3} Ionization potentials of transition-metal atoms.}
\end{figure}

Fig. \ref{fig3} shows the calculated ionization potentials (IPs) of the first-row transition-metal (TM) atoms (Sc-Zn). The IPs were calculated by the energy difference between the positive ions and the neutral atoms. The data sets for these graphs can be found in Table I of the SM. The inspection of Fig. \ref{fig3} reveals that calculated GNOF IPs are close to the CCSD(T) values, although they deviate from the latter and are closer to the experimental values for early TMs (Sc-V). Note that the MAE with respect to the experiment is similar for both methods, 7.9 and 7.3 kcal/mol, respectively, which
is an outstanding result considering the size of the basis sets employed.

\begin{table}[H]
\caption{\label{table2} Mean absolute differences (mHartrees) with respect to CCSD(T) values for total electronic energies.}
\smallskip{}
\centering{}%
\begin{tabular}{ccccc||c||c||c||c}
\hline 
\multicolumn{2}{c}{Molecules} & $\enskip$MP2$\enskip$ & CCSD & \multicolumn{5}{c}{GNOF}\tabularnewline
\hline 
singlets    & (30)$\enskip$ & 30.58 & 18.67    & \multicolumn{5}{c}{7.66}\tabularnewline
multiplets  & (25)$\enskip$ & 28.43 & $\,$9.39 & \multicolumn{5}{c}{7.83}\tabularnewline
\hline 
\end{tabular}
\end{table}

Table \ref{table2} shows the mean absolute differences with respect to the CCSD(T) values for electronic energies of 55 selected molecules in different spin states calculated at the experimental geometries using the MP2, CCSD and GNOF methods. The energies of the 30 singlets and 25 multiplets considered can be found in Tables II and III of the SM, respectively. For the whole set, the average differences in the MP2, CCSD and GNOF energies from CCSD(T) are 29.7, 14.4 and 7.7 mHartree, respectively. These differences reveal the good performance of GNOF for molecular energies, and no important differences are observed in relation to the spin of the system.

\bigskip{}\bigskip{}

\begin{figure}[th]
\centering{}\includegraphics[scale=0.4]{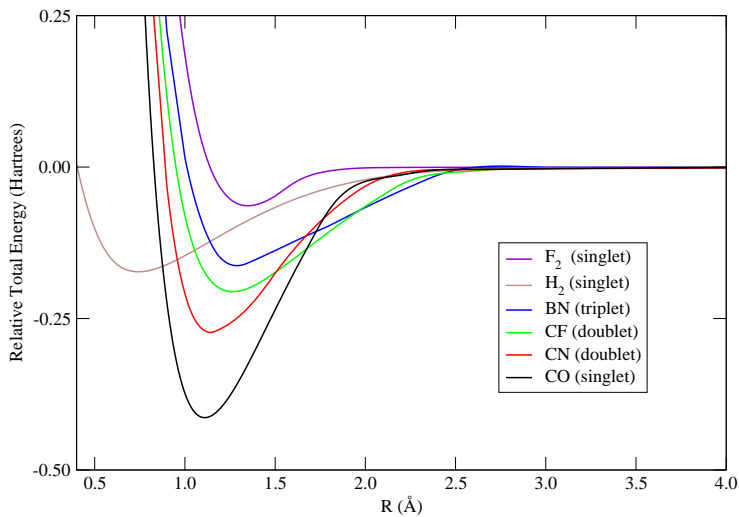}\caption{\label{fig:PECs}Potential Energy Curves.}
\end{figure}

\begin{table}[h]
\caption{\label{ReDeWe}Comparison of R$_{e}$ ($\textrm{\AA}$), D$_{e}$ (kcal/mol), and $\omega_{e}$ (cm$^{-1}$) with the experimental values.}
\bigskip{}
\centering{}%
\begin{tabular}{lcccccccccccc}
\hline 
Mol & Mul & R$_{e}$ &  & $\mathrm{R_{\mathit{e}}^{\mathit{exp}}}$ & \enskip{} & D$_{e}$ &  & $\mathrm{D_{\mathit{e}}^{\mathit{exp}}}$ & \enskip{} & $\omega_{e}$ &  & $\mathrm{\omega_{\mathit{e}}^{\mathit{exp}}}$\tabularnewline
\hline 
F$_{2}$  & 1 & 1.35 &  & 1.41 &  & \enskip{}40.9 &  & \enskip{}39.2 &  & 1212 &  & \enskip{}917 \tabularnewline
H$_{2}$  & 1 & 0.74 &  & 0.74 &  &         108.6 &  &         109.5 &  & 4404 &  &         4401 \tabularnewline
BN       & 3 & 1.29 &  & 1.32 &  &         102.3 &  &        94-133 &  & 1851 &  &         1515 \tabularnewline
CN       & 2 & 1.14 &  & 1.17 &  &         171.6 &  &         177.4 &  & 2344 &  &         2069 \tabularnewline
CF       & 2 & 1.26 &  & 1.27 &  &         129.0 &  &         128.7 &  & 1238 &  &         1308 \tabularnewline
CO       & 1 & 1.11 &  & 1.13 &  &         259.6 &  &         259.3 &  & 2391 &  &         2170 \tabularnewline
\hline 
\end{tabular}
\end{table}

The performance of GNOF has also been tested in the dissociation of diatomic molecules. Representative potential energy curves (PECs) of six dimers with different values of total spin are depicted in Fig. \ref{fig:PECs}. The zero energy for each curve has been set at 10 Å. At the equilibrium, these dimers comprise different types of bonds, from single to triple bonds. However, in all cases the correct dissociation limit implies an homolytic cleavage of the bonds with high degree of degeneracy effects depending on the multiplicity of the dissociated atoms (see Table \ref{table1}). It is well known that density functionals tend to dissociate to atoms with spurious fractional charge \citep{Perdew1982}, especially in heteronuclear species. In contrast, GNOF produces correct PECs with dissociation limits that have integer numbers of electrons in the dissociated atoms in all cases. Find illustrative comparisons between GNOF and CASPT2 methods for CF and CO dimers in Figs. 2 and 3 of the SM.

In Table \ref{ReDeWe}, selected electronic properties, including equilibrium distances (R$_{e}$), dissociation energies (D$_{e}$), and harmonic vibrational frequencies ($\omega_{e}$) can be found. In general, it can be seen that GNOF underestimates the equilibrium distances and overestimates the frequencies, while giving a better agreement for the binding energies. The quality of the electronic
structure description in the equilibrium region can be seen in CO, for which GNOF predicts a dipole moment of 0.107$\,D$ with the correct sign, in good agreement with the experimental value of 0.112$\,D$,
contrary to HF or CASSCF results.

The performance of GNOF has also been investigated in the treatment of near-degeneracy effects in reactions in which diradicals are formed. A paradigmatic case is the ethylene torsion, where a full degeneracy of the $\pi$ orbital system is observed for $90\lyxmathsym{\textdegree}$ torsion angle. In terms of relative energies, single-reference methods greatly overestimates the barrier
height, which decreases when near-degeneracy effects are considered. GNOF predicts a barrier of 3.19 eV using the cc-pVDZ basis set \citep{Pritchard2019}, in outstanding agreement with the result of the SF-CIS(D) method \citep{Casanova2008}. Furthermore, the GNOF ONs at a $90\lyxmathsym{\textdegree}$ torsion angle for the valence $\pi$ orbitals are equal to 1.000, corresponding to the correct description of these fully degenerate orbitals.

\smallskip{}

\textbf{Acknowledgments:} Financial support comes from MCIU/AEI/FEDER, 
UE(PGC2018-097529-B-100) and Eusko Jaurlaritza (Ref. IT1254-19). 
The author thanks for the support provided by IZO-SGI SGIker of UPV/EHU 
and European funding (ERDF and ESF)

\bigskip{}

\bigskip{}\bigskip{}

\begin{center}
\textbf{Supplementary Material}
\end{center}

\setcounter{figure}{0}
\setcounter{table}{0}

\begin{center}
\begin{figure}[H]
\begin{centering}
\caption{\label{fig1} Illustrative example of splitting of the orbital space $\Omega$ into subspaces: $\Omega=\Omega_{\mathrm{I}}\oplus\Omega_{\mathrm{II}}=\Omega^{a}\oplus\Omega^{b},\,\Omega_{\mathrm{II}}=\Omega_{\mathrm{II}}^{a}\oplus\Omega_{\mathrm{II}}^{b}$. $\Omega^{a}$ ($\Omega^{b}$) denotes the subspace composed of orbitals above (below) the level $\mathrm{N}_{\Omega}$, that is, $\Omega^{a}\equiv p>\mathrm{N}_{\Omega}$ ($\Omega^{b}\equiv p\protect\leq\mathrm{N}_{\Omega}$). Similarly, $\Omega_{\mathrm{II}}^{b}\equiv p\protect\leq\mathrm{N_{II}}/2$ and $\Omega_{\mathrm{II}}^{a}\equiv p>\mathrm{N}_{\Omega}$. In this example, $S=1$ (triplet) and $\mathrm{N_{I}}=2$, so two orbitals make up the subspace $\Omega_{\mathrm{I}}$, whereas fourteen electrons ($\mathrm{N_{II}}=14$)
distributed in seven subspaces $\left\{ \Omega_{1},\Omega_{2},...,\Omega_{7}\right\} $ make up the subspace $\Omega_{\mathrm{II}}$. Note that $\mathrm{N}_{g}=2$ for all subspaces $\Omega{}_{g}\in\Omega_{\mathrm{II}}$, and $\mathrm{N}_{\Omega}=\mathrm{N_{II}}/2+\mathrm{N_{I}}=9$. The arrows depict the values of the ensemble occupancies, alpha ($\downarrow$) or beta ($\uparrow$), in each orbital.}
\par\end{centering}
\centering{}\includegraphics[scale=0.4]{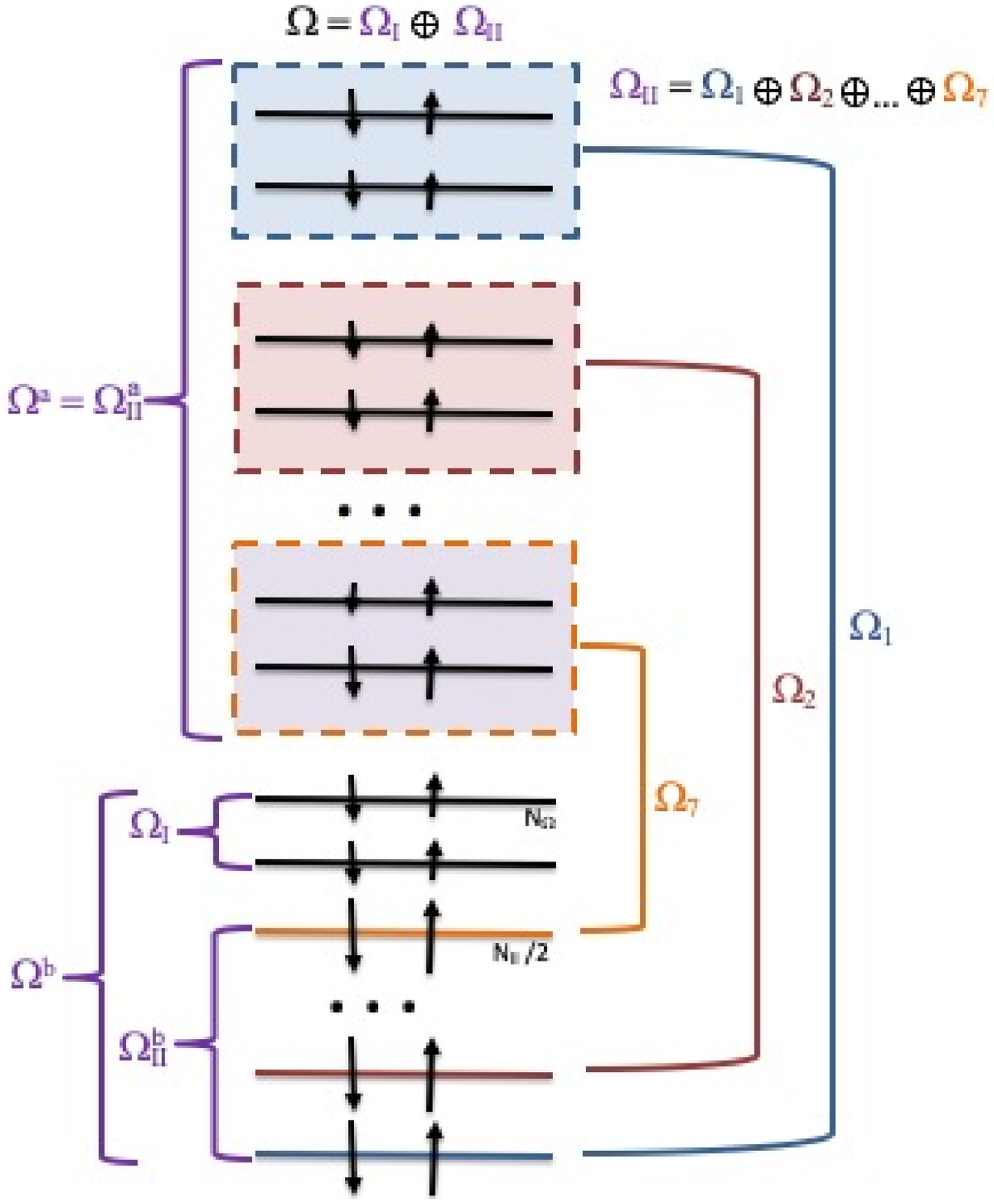}
\end{figure}
\par\end{center}

\textbf {GNOF for Singlet States:}

\begin{equation}
E = E^{intra} + E^{inter} \tag{A}
\end{equation}

\begin{equation}
\begin{array}{c}
E^{intra} = \sum\limits _{g=1}^{\mathrm{N}_{\Omega}}E_{g} \quad,\qquad E_{g}=\sum\limits _{p\in\Omega_{g}}n_{p}\left(2H_{pp}+L_{pp}\right)+{\displaystyle \sum\limits _{p,q\in\Omega_{g},p\neq q}}\Pi\left(n_{q},n_{p}\right)L_{pq}\tag{B}
\end{array}
\end{equation}

\begin{equation}
\begin{array}{c}
E^{inter}=\sum\limits _{p,q=1}^{\mathrm{N}_{B}}\,'\,\left\{ n_{q}n_{p}\left(2J_{pq}-K_{pq}\right)+\left[n_{q}^{d}n_{p}^{d} + \;\Pi\left(n_{q}^{d},n_{p}^{d}\right)-\Phi_{q}\Phi_{p}\right]\left(1-\delta_{q\Omega^{b}}\delta_{p\Omega^{b}}\right)L_{pq}\right\}\tag{C}
\end{array}
\end{equation}

\noindent where

\begin{equation}
\Phi_{p}=\sqrt{n_{p}(1-n_{p})} \quad,\qquad \Pi\left(n_{q},n_{p}\right)=\sqrt{n_{q}n_{p}}\left(\delta_{q\Omega^{a}}\delta_{p\Omega^{a}}-\delta_{qg}-\delta_{pg}\right)\tag{D}
\end{equation}

\bigskip

\noindent $H_{pp}$ denotes the diagonal one-electron matrix elements of the kinetic energy and external potential operators. $J_{pq}=\left\langle pq|pq\right\rangle $, $K_{pq}=\left\langle pq|qp\right\rangle $, and $L_{pq}=\left\langle pp|qq\right\rangle $ are the Coulomb, exchange, and exchange-time-inversion integrals, respectively. $\Omega^{b}$ denotes the subspace composed of orbitals below the level $\mathrm{N}_{\Omega}$ ($p\leq\mathrm{N}_{\Omega}$), whereas $\Omega^{a}$ denotes the subspace composed of orbitals above the level $\mathrm{N}_{\Omega}$ ($p>\mathrm{N}_{\Omega}$). In the last summation, the prime indicates that only the inter-subspace terms are taking into account ($p\in\Omega{}_{f},q\in\Omega{}_{g},f\neq g$).

\bigskip

The dynamic part of the occupation number $n_{p}$ is defined as

\begin{equation}
n_{p}^{d}=n_{p}\cdot e^{-\left(\dfrac{h_{g}}{h_{c}}\right)^{2}},\,p\in\Omega_{g}\,,\enskip g=1,2,...,{\displaystyle \mathrm{N}_{\Omega}}\label{dyn-on}
\end{equation}

\begin{table}[b]
\caption{Ionizations potentials (kcal/mol) calculated as $\mathrm{IP}=E(\mathrm{X^{+}})-E(\mathrm{X})$ using the cc-pVTZ basis set. CCSD(T) and experimental values taken from Table IV of Ref. \citep{Balabanov2005}. MAE corresponds to the mean absolute error with respect to experiment.}
\bigskip

\centering{}
\begin{tabular}{clccccccc}
\hline 
\multirow{1}{*}{}
Atom &  & X &  & X$^{+}$ &  & GNOF & CCSD(T) & EXP \tabularnewline
\hline 
Sc &  & $^{2}$D &  & $^{3}$D &  & 156.9 & 146.8 & 151.3 \tabularnewline
Ti &  & $^{3}$F &  & $^{4}$F &  & 161.9 & 152.8 & 157.5 \tabularnewline
V  &  & $^{4}$F &  & $^{5}$D &  & 161.6 & 149.4 & 155.2 \tabularnewline
Cr &  & $^{7}$S &  & $^{6}$S &  & 142.4 & 147.9 & 156.0 \tabularnewline
Mn &  & $^{6}$S &  & $^{7}$S &  & 167.5 & 166.0 & 171.4 \tabularnewline
Fe &  & $^{5}$D &  & $^{6}$D &  & 180.2 & 176.2 & 182.3 \tabularnewline
Co &  & $^{4}$F &  & $^{3}$F &  & 175.5 & 173.3 & 181.5 \tabularnewline
Ni &  & $^{3}$F &  & $^{2}$D &  & 161.4 & 164.5 & 175.1 \tabularnewline
Cu &  & $^{2}$S &  & $^{1}$S &  & 168.2 & 167.0 & 178.2 \tabularnewline
Zn &  & $^{1}$S &  & $^{2}$S &  & 203.7 & 208.0 & 216.6 \tabularnewline
\hline 
\multicolumn{6}{c}{$\quad\qquad$MAE} & 7.9 & 7.3 & -\tabularnewline
\hline 
\end{tabular}
\end{table}

\newpage

\begin{figure}[th]
\centering{}\includegraphics[scale=0.50]{co}\caption{\label{fig:PEC1}Potential energy curves for the singlet ground state of the CO molecule obtained with GNOF and CASPT2 (including 6 active electrons on 6 active orbitals) methods using the cc-pVDZ basis set. The author appreciates the CASPT2 values provided by Dr. M. Rodriguez-Mayorga.}
\end{figure}

\begin{figure}[th]
\centering{}\includegraphics[scale=0.50]{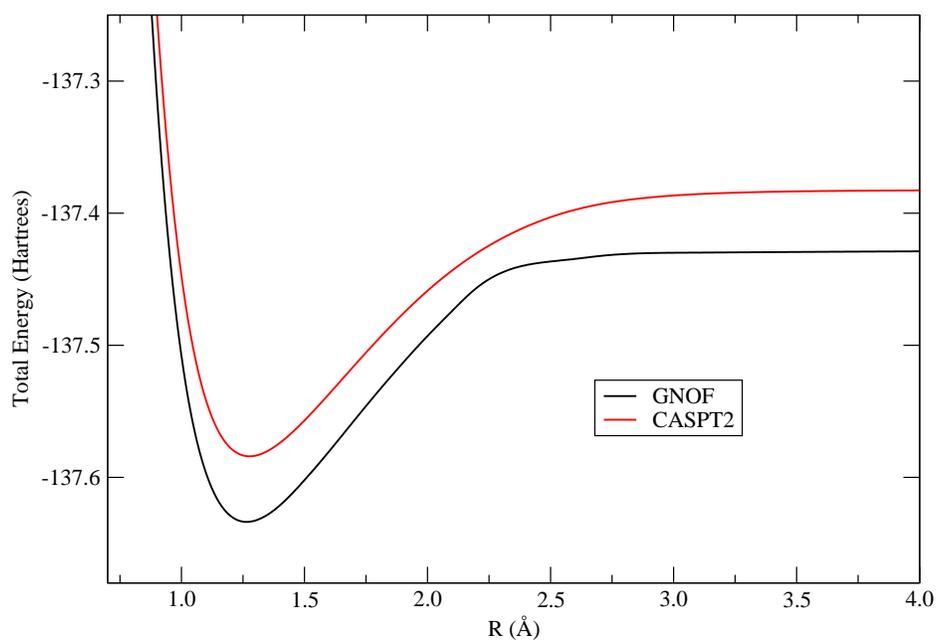}\caption{\label{fig:PEC2}Potential energy curves for the doublet ground state of the CF molecule obtained with GNOF and CASPT2 (including 9 active electrons on 9 active orbitals) methods using the cc-pVTZ basis set. The CASPT2 values were taken from Ref. \citep{Quintero-Monsebaiz2021}.}
\end{figure}

\begin{table}[H]
\caption{Comparison of total energies, in Hartrees, calculated at the MP2, GNOF, and CCSD(T) levels of theory for 30 singlet species. Calculations were performed using the experimental geometries of Ref. \citep{NIST} and the cc-pVTZ basis set \citep{Pritchard2019}. MAE corresponds to the mean absolute error with respect to the CCSD(T) values.\bigskip{}
}
\centering{}
\begin{tabular}{ccccc}
\hline 
 & \multicolumn{2}{c}{} &  & \tabularnewline
Molecule & $\qquad\qquad$MP2$\qquad$$\qquad$ & $\qquad$CCSD$\qquad$ & $\qquad$GNOF$\qquad$ & $\qquad$CCSD(T)$\qquad$\tabularnewline
 &  &  &  & \tabularnewline
\hline 
 &  &  &  & \tabularnewline
H$_{2}$ & $\,$$\,$-1.16477  & $\,$$\,$-1.17246 & $\,$$\,$-1.17245 & $\,$$\,$-1.17246\tabularnewline
Li$_{2}$ & $\,$-14.92088  & $\,$-14.93580 & $\,$-14.93403 & $\,$-14.93614\tabularnewline
LiF & -107.29224 & -107.29141 & -107.29480 & -107.29944\tabularnewline
CO & -113.16682 & -113.16997 & -113.18076 & -113.18735\tabularnewline
N$_{2}$ & -109.39143  & -109.38967 & -109.39820 & -109.40874\tabularnewline
F$_{2}$ & -199.31048 & -199.31389 & -199.33563 & -199.33236\tabularnewline
HCN & $\,$-93.25775 & $\,$-93.26678 & $\,$-93.27677 & $\,$-93.28278\tabularnewline
CO$_{2}$ & -188.36075  & -188.35149 & -188.36420 & -188.38074\tabularnewline
BF$_{3}$ & -324.23564  & -324.23634 & -324.26896 & -324.26147\tabularnewline
HF & -100.34838 & -100.35043 & -100.36077 & -100.35698\tabularnewline
NH$_{3}$ & $\,$-56.47205  & $\,$-56.48460 & $\,$-56.48734 & $\,$-56.49251\tabularnewline
H$_{2}$O & $\,$-76.33668  & $\,$-76.34239 & $\,$-76.34500 & $\,$-76.35029\tabularnewline
CH$_{4}$ & $\,$-40.43238 & $\,$-40.45308 & $\,$-40.45533 & $\,$-40.45960\tabularnewline
C$_{2}$H$_{2}$ & $\,$-77.19552 & $\,$-77.20852 & $\,$-77.20895 & $\,$-77.22541\tabularnewline
C$_{2}$H$_{4}$ & $\,$-78.43744 & $\,$-78.46281 & $\,$-78.46934 & $\,$-78.47847 \tabularnewline
N$_{2}$H$_{4}$ & -111.69571 & -111.71397 & -111.72556 & -111.73123\tabularnewline
C$_{2}$H$_{6}$ & $\,$-79.67171 & $\,$-79.70386 & $\,$-79.71166 & $\,$-79.71789\tabularnewline
H$_{2}$O$_{2}$ & -151.36567 & -151.37226 & -151.39239 & -151.39166\tabularnewline
H$_{2}$CO & -114.34175 & -114.35216 & -114.36809 & -114.36928\tabularnewline
HCOOH & -189.51455 & -189.51959 & -189.54192  & -189.54659\tabularnewline
CH$_{2}$CO  & -152.37891 & -152.38801 & -152.40087  & -152.41506\tabularnewline
C$_{2}$FH$_{3}$  & -177.58430 & -177.60409 & -177.62610 & -177.62758\tabularnewline
C$_{2}$H$_{4}$O & -153.55941 & -153. 57865 & -153.59681 & -153.60360\tabularnewline
C$_{2}$H$_{5}$N & -133.70022 & -133.72502 & -133.73644 & -133.74992\tabularnewline
C$_{2}$H$_{2}$O$_{2}$  & -227.51149 & -227.52057 & -227.54481 & -227.55734\tabularnewline
CH$_{3}$CN & -132.39931  & -132.41060 & -132.41060 & -132.43869\tabularnewline
CH$_{3}$NH$_{2}$ & $\,$-95.69653 & $\,$-95.72131 & $\,$-95.73376 & $\,$-95.73676\tabularnewline
CH$_{3}$NO$_{2}$  & -244.69556 & -244.69536 & -244.71682 & -244.73859\tabularnewline
CH$_{3}$OCH$_{3}$  & -154.78170 & -154.81254 & -154.83262 & -154.83576\tabularnewline
CH$_{3}$CH$_{2}$OH  & -154.80194 & -154.83142 & -154.84906 & -154.85458\tabularnewline
 &  &  &  & \tabularnewline
\hline 
 &  &  &  & \tabularnewline
MAE & $\:$0.03058 & $\:$0.018674 & $\:$0.00766 & -\tabularnewline
\end{tabular}
\end{table}

\begin{table}[H]
\caption{Comparison of total energies, in Hartrees, calculated at the MP2, GNOF, and CCSD(T) levels of theory for 25 multiplet species. Calculations were performed using the experimental geometries of Ref. \citep{NIST} and the cc-pVTZ basis set \citep{Pritchard2019}. For cations, the experimental geometry of the neutral species was used. MAE corresponds to the mean absolute error with respect to the CCSD(T) values.\bigskip{}
}

\centering{}%
\begin{tabular}{cccccc}
\hline 
 &  & \multicolumn{2}{c}{} &  & \tabularnewline
Molecule & $\quad$Mul$\quad$ & $\qquad\qquad$MP2$\qquad$$\qquad$ & $\qquad$CCSD$\qquad$ & $\qquad$GNOF$\qquad$ & $\qquad$CCSD(T)$\qquad$\tabularnewline
 &  &  &  &  & \tabularnewline
\hline 
 &  &  &  &  & \tabularnewline
B$_{2}$ & 3 & $\,$-49.27651 & $\,$-49.30265 & $\,$-49.31599 & $\,$-49.32052\tabularnewline
BN & 3 & $\,$-79.26194 & $\,$-79.28174 & $\,$-79.27990 & $\,$-79.29500\tabularnewline
CN & 2 & $\,$-92.54012 & $\,$-92.57821 & $\,$-92.56554 & $\,$-92.59725\tabularnewline
NO & 2 & -129.72508 & -129.73125 & -129.72614 & -129.75021\tabularnewline
CF & 2 & -137.60791 & -137.62228 & -137.63002 & -137.63567\tabularnewline
NF & 3 & -154.25367 & -154.26800 & -154.26870 & -154.28207\tabularnewline
$\mathrm{Li_{2}^{+}}$ & 2 & $\,$-14.73922 & $\,$-14.74404 & $\,$-14.74167 & $\,$-14.74407\tabularnewline
LiF$^{+}$ & 2 & -106.86260 & -106.87655 & -106.88263 & -106.88046\tabularnewline
BeH & 2 & $\,$-15.20111  & $\,$-15.21256 & $\,$-15.21649 & $\,$-15.21318\tabularnewline
CH & 2 & $\,$-38.39552  & $\,$-38.42014 & $\,$-38.42753 & $\,$-38.42365\tabularnewline
CH$_{2}$ & 3 & $\,$-39.07065 & $\,$-39.09023 & $\,$-39.09155 & $\,$-39.09381\tabularnewline
CH$_{3}$ & 2 & $\,$-39.75360 & $\,$-39.77468 & $\,$-39.77716 & $\,$-39.77962\tabularnewline
$\mathrm{CH}_{4}^{+}$ & 2 & $\,$-39.90211 & $\,$-39.92579 & $\,$-39.93310 & $\,$-39.93130\tabularnewline
NH & 3 & $\,$-55.13293 & $\,$-55.15196 & $\,$-55.15363 & $\,$-55.15593\tabularnewline
NH$_{2}$ & 2 & $\,$-55.78739 & $\,$-55.80484 & $\,$-55.80867 & $\,$-55.81070\tabularnewline
OH$^{+}$ & 3 & $\,$-75.16360  & $\,$-75.18188 & $\,$-75.18683 & $\,$-75.18510\tabularnewline
OH & 2 & $\,$-75.63531 & $\,$-75.64899 & $\,$-75.65656 & $\,$-75.65427\tabularnewline
OOH & 2 & -150.71333 & -150.72932 & -150.73734 & -150.74695\tabularnewline
HCO & 2 & -113.69156 & -113.69792 & -113.69307 & -113.71590\tabularnewline
HCN$^{+}$ & 2 & $\,$-92.74287  & $\,$-92.76755 & $\,$-92.76268 & $\,$-92.78106\tabularnewline
H$_{2}$O$^{+}$ & 2 & $\,$-75.87013  & $\,$-75.88556 & $\,$-75.89391 & $\,$-75.89045\tabularnewline
$\mathrm{NH}_{3}^{+}$ & 2 & $\,$-56.07287  & $\,$-56.09136 & $\,$-56.09831 & $\,$-56.09634\tabularnewline
$\mathrm{C_{2}H_{2}^{+}}$ & 2 & $\,$-76.76931  & $\,$-76.79418 & $\,$-76.80360 & $\,$-76.80539\tabularnewline
C$_{2}$H$_{3}$ & 2 & $\,$-77.740316 & $\,$-77.77558 & $\,$-77.77546 & $\,$-77.78980\tabularnewline
CH$_{3}$O & 2 & -114.86763 & -114.89605 & -114.90702 & -114.90933\tabularnewline
 &  &  &  &  & \tabularnewline
\hline 
 &  &  &  &  & \tabularnewline
\multicolumn{2}{c}{MAE} & $\:$0.02843 & $\:$0.009389 & $\:$0.00783 & -\tabularnewline
\end{tabular}
\end{table}


\begin{thebibliography}{43}
\expandafter\ifx\csname natexlab\endcsname\relax\def\natexlab#1{#1}\fi
\expandafter\ifx\csname bibnamefont\endcsname\relax
  \def\bibnamefont#1{#1}\fi
\expandafter\ifx\csname bibfnamefont\endcsname\relax
  \def\bibfnamefont#1{#1}\fi
\expandafter\ifx\csname citenamefont\endcsname\relax
  \def\citenamefont#1{#1}\fi
\expandafter\ifx\csname url\endcsname\relax
  \def\url#1{\texttt{#1}}\fi
\expandafter\ifx\csname urlprefix\endcsname\relax\def\urlprefix{URL }\fi
\providecommand{\bibinfo}[2]{#2}
\providecommand{\eprint}[2][]{\url{#2}}

\bibitem[{\citenamefont{Gilbert}(1975)}]{Gilbert1975}
\bibinfo{author}{\bibfnamefont{T.~L.} \bibnamefont{Gilbert}},
  \bibinfo{journal}{Phys. Rev. B} \textbf{\bibinfo{volume}{12}},
  \bibinfo{pages}{2111} (\bibinfo{year}{1975}).

\bibitem[{\citenamefont{Valone}(1980)}]{Valone1980}
\bibinfo{author}{\bibfnamefont{S.~M.} \bibnamefont{Valone}},
  \bibinfo{journal}{J. Chem. Phys.} \textbf{\bibinfo{volume}{73}},
  \bibinfo{pages}{1344} (\bibinfo{year}{1980}).

\bibitem[{\citenamefont{Levy}(1979)}]{Levy1979}
\bibinfo{author}{\bibfnamefont{M.}~\bibnamefont{Levy}}, \bibinfo{journal}{Proc.
  Natl. Acad. Sci. USA} \textbf{\bibinfo{volume}{76}}, \bibinfo{pages}{6062}
  (\bibinfo{year}{1979}).

\bibitem[{\citenamefont{Coleman}(1963)}]{Coleman1963}
\bibinfo{author}{\bibfnamefont{A.~J.} \bibnamefont{Coleman}},
  \bibinfo{journal}{Rev. Mod. Phys.} \textbf{\bibinfo{volume}{35}},
  \bibinfo{pages}{668} (\bibinfo{year}{1963}).

\bibitem[{\citenamefont{Lude\~{n}a et~al.}(2013)\citenamefont{Lude\~{n}a,
  Torres, and Costa}}]{Ludena2013}
\bibinfo{author}{\bibfnamefont{E.~V.} \bibnamefont{Lude\~{n}a}},
  \bibinfo{author}{\bibfnamefont{F.~J.} \bibnamefont{Torres}},
  \bibnamefont{and} \bibinfo{author}{\bibfnamefont{C.}~\bibnamefont{Costa}},
  \bibinfo{journal}{J. Mod. Phys.} \textbf{\bibinfo{volume}{04}},
  \bibinfo{pages}{391} (\bibinfo{year}{2013}).

\bibitem[{\citenamefont{Piris}(2018{\natexlab{a}})}]{Piris2018}
\bibinfo{author}{\bibfnamefont{M.}~\bibnamefont{Piris}}, in
  \emph{\bibinfo{booktitle}{Many-body approaches at different scales: a tribute
  to N. H. March on the occasion of his 90th birthday}}, edited by
  \bibinfo{editor}{\bibfnamefont{G.~G.~N.} \bibnamefont{Angilella}}
  \bibnamefont{and} \bibinfo{editor}{\bibfnamefont{C.}~\bibnamefont{Amovilli}}
  (\bibinfo{publisher}{Springer}, \bibinfo{address}{New York},
  \bibinfo{year}{2018}{\natexlab{a}}), chap.~\bibinfo{chapter}{22}, pp.
  \bibinfo{pages}{283--300}.

\bibitem[{\citenamefont{Mazziotti}(2012{\natexlab{a}})}]{Mazziotti2012}
\bibinfo{author}{\bibfnamefont{D.~A.} \bibnamefont{Mazziotti}},
  \bibinfo{journal}{Phys. Rev. Lett.} \textbf{\bibinfo{volume}{108}},
  \bibinfo{pages}{263002} (\bibinfo{year}{2012}{\natexlab{a}}).

\bibitem[{\citenamefont{Mazziotti}(2012{\natexlab{b}})}]{Mazziotti2012b}
\bibinfo{author}{\bibfnamefont{D.~A.} \bibnamefont{Mazziotti}},
  \bibinfo{journal}{Phys. Rev. A} \textbf{\bibinfo{volume}{85}},
  \bibinfo{pages}{062507} (\bibinfo{year}{2012}{\natexlab{b}}).

\bibitem[{\citenamefont{Piris}(2007)}]{Piris2007}
\bibinfo{author}{\bibfnamefont{M.}~\bibnamefont{Piris}}, in
  \emph{\bibinfo{booktitle}{Reduced-Density-Matrix Mechanics: with applications
  to many-electron atoms and molecules}}, edited by
  \bibinfo{editor}{\bibfnamefont{D.~A.} \bibnamefont{Mazziotti}}
  (\bibinfo{publisher}{John Wiley and Sons}, \bibinfo{address}{Hoboken, New
  Jersey, USA}, \bibinfo{year}{2007}), chap.~\bibinfo{chapter}{14}, pp.
  \bibinfo{pages}{387--427}.

\bibitem[{\citenamefont{Piris and Ugalde}(2014)}]{Piris2014a}
\bibinfo{author}{\bibfnamefont{M.}~\bibnamefont{Piris}} \bibnamefont{and}
  \bibinfo{author}{\bibfnamefont{J.~M.} \bibnamefont{Ugalde}},
  \bibinfo{journal}{Int. J. Quantum Chem.} \textbf{\bibinfo{volume}{114}},
  \bibinfo{pages}{1169} (\bibinfo{year}{2014}).

\bibitem[{\citenamefont{Pernal and Giesbertz}(2016)}]{Pernal2016}
\bibinfo{author}{\bibfnamefont{K.}~\bibnamefont{Pernal}} \bibnamefont{and}
  \bibinfo{author}{\bibfnamefont{K.~J.~H.} \bibnamefont{Giesbertz}},
  \bibinfo{journal}{Top Curr Chem} \textbf{\bibinfo{volume}{368}},
  \bibinfo{pages}{125} (\bibinfo{year}{2016}).

\bibitem[{\citenamefont{Schade et~al.}(2017)\citenamefont{Schade, Kamil, and
  Bl{\"{o}}chl}}]{Schade2017}
\bibinfo{author}{\bibfnamefont{R.}~\bibnamefont{Schade}},
  \bibinfo{author}{\bibfnamefont{E.}~\bibnamefont{Kamil}}, \bibnamefont{and}
  \bibinfo{author}{\bibfnamefont{P.}~\bibnamefont{Bl{\"{o}}chl}},
  \bibinfo{journal}{Eur. Phys. J. Spec. Top.} \textbf{\bibinfo{volume}{226}},
  \bibinfo{pages}{2677} (\bibinfo{year}{2017}).

\bibitem[{\citenamefont{Mitxelena et~al.}(2019)\citenamefont{Mitxelena, Piris,
  and Ugalde}}]{Mitxelena2019}
\bibinfo{author}{\bibfnamefont{I.}~\bibnamefont{Mitxelena}},
  \bibinfo{author}{\bibfnamefont{M.}~\bibnamefont{Piris}}, \bibnamefont{and}
  \bibinfo{author}{\bibfnamefont{J.~M.} \bibnamefont{Ugalde}}, in
  \emph{\bibinfo{booktitle}{State Art Mol. Electron. Struct. Comput. Correl.
  Methods, Basis Sets More}}, edited by
  \bibinfo{editor}{\bibfnamefont{P.}~\bibnamefont{Hoggan}} \bibnamefont{and}
  \bibinfo{editor}{\bibfnamefont{U.}~\bibnamefont{Ancarani}}
  (\bibinfo{publisher}{Academic Press}, \bibinfo{year}{2019}),
  chap.~\bibinfo{chapter}{7}, pp. \bibinfo{pages}{155--177}.

\bibitem[{\citenamefont{Muller}(1984)}]{Muller1984}
\bibinfo{author}{\bibfnamefont{A.~M.~K.} \bibnamefont{Muller}},
  \bibinfo{journal}{Phys. Lett.} \textbf{\bibinfo{volume}{105A}},
  \bibinfo{pages}{446} (\bibinfo{year}{1984}).

\bibitem[{\citenamefont{Goedecker and Umrigar}(1998)}]{Goedecker1998}
\bibinfo{author}{\bibfnamefont{S.}~\bibnamefont{Goedecker}} \bibnamefont{and}
  \bibinfo{author}{\bibfnamefont{C.}~\bibnamefont{Umrigar}},
  \bibinfo{journal}{Phys. Rev. Lett.} \textbf{\bibinfo{volume}{81}},
  \bibinfo{pages}{866} (\bibinfo{year}{1998}).

\bibitem[{\citenamefont{Csanyi and Arias}(2000)}]{Csanyi2000}
\bibinfo{author}{\bibfnamefont{G.}~\bibnamefont{Csanyi}} \bibnamefont{and}
  \bibinfo{author}{\bibfnamefont{T.~A.} \bibnamefont{Arias}},
  \bibinfo{journal}{Phys. Rev. B} \textbf{\bibinfo{volume}{61}},
  \bibinfo{pages}{7348} (\bibinfo{year}{2000}).

\bibitem[{\citenamefont{Gritsenko et~al.}(2005)\citenamefont{Gritsenko, Pernal,
  and Baerends}}]{Gritsenko2005}
\bibinfo{author}{\bibfnamefont{O.~V.} \bibnamefont{Gritsenko}},
  \bibinfo{author}{\bibfnamefont{K.}~\bibnamefont{Pernal}}, \bibnamefont{and}
  \bibinfo{author}{\bibfnamefont{E.~J.} \bibnamefont{Baerends}},
  \bibinfo{journal}{J. Chem. Phys.} \textbf{\bibinfo{volume}{122}},
  \bibinfo{pages}{204102} (\bibinfo{year}{2005}).

\bibitem[{\citenamefont{Sharma et~al.}(2008)\citenamefont{Sharma, Dewhurst,
  Lathiotakis, and Gross}}]{Sharma2008}
\bibinfo{author}{\bibfnamefont{S.}~\bibnamefont{Sharma}},
  \bibinfo{author}{\bibfnamefont{J.~K.} \bibnamefont{Dewhurst}},
  \bibinfo{author}{\bibfnamefont{N.~N.} \bibnamefont{Lathiotakis}},
  \bibnamefont{and} \bibinfo{author}{\bibfnamefont{E.~K.~U.}
  \bibnamefont{Gross}}, \bibinfo{journal}{Phys. Rev. B}
  \textbf{\bibinfo{volume}{78}}, \bibinfo{pages}{201103(R)}
  (\bibinfo{year}{2008}).

\bibitem[{\citenamefont{Marques and Lathiotakis}(2008)}]{Marques2008}
\bibinfo{author}{\bibfnamefont{M.~A.~L.} \bibnamefont{Marques}}
  \bibnamefont{and} \bibinfo{author}{\bibfnamefont{N.~N.}
  \bibnamefont{Lathiotakis}}, \bibinfo{journal}{Phys. Rev. A}
  \textbf{\bibinfo{volume}{77}}, \bibinfo{pages}{032509}
  (\bibinfo{year}{2008}).

\bibitem[{\citenamefont{Rohr et~al.}(2008)\citenamefont{Rohr, Pernal,
  Gritsenko, and Baerends}}]{Rohr2008}
\bibinfo{author}{\bibfnamefont{D.~R.} \bibnamefont{Rohr}},
  \bibinfo{author}{\bibfnamefont{K.}~\bibnamefont{Pernal}},
  \bibinfo{author}{\bibfnamefont{O.~V.} \bibnamefont{Gritsenko}},
  \bibnamefont{and} \bibinfo{author}{\bibfnamefont{E.~J.}
  \bibnamefont{Baerends}}, \bibinfo{journal}{J. Chem. Phys.}
  \textbf{\bibinfo{volume}{129}}, \bibinfo{pages}{164105}
  (\bibinfo{year}{2008}).

\bibitem[{\citenamefont{Rodr{\'{i}}guez-Mayorga
  et~al.}(2017)\citenamefont{Rodr{\'{i}}guez-Mayorga, Ramos-Cordoba, Via-Nadal,
  Piris, and Matito}}]{Rodriguez-Mayorga2017}
\bibinfo{author}{\bibfnamefont{M.}~\bibnamefont{Rodr{\'{i}}guez-Mayorga}},
  \bibinfo{author}{\bibfnamefont{E.}~\bibnamefont{Ramos-Cordoba}},
  \bibinfo{author}{\bibfnamefont{M.}~\bibnamefont{Via-Nadal}},
  \bibinfo{author}{\bibfnamefont{M.}~\bibnamefont{Piris}}, \bibnamefont{and}
  \bibinfo{author}{\bibfnamefont{E.}~\bibnamefont{Matito}},
  \bibinfo{journal}{Phys. Chem. Chem. Phys.} \textbf{\bibinfo{volume}{19}},
  \bibinfo{pages}{24029} (\bibinfo{year}{2017}).

\bibitem[{\citenamefont{Piris}(2013{\natexlab{a}})}]{Piris2013b}
\bibinfo{author}{\bibfnamefont{M.}~\bibnamefont{Piris}}, \bibinfo{journal}{Int.
  J. Quantum Chem.} \textbf{\bibinfo{volume}{113}}, \bibinfo{pages}{620}
  (\bibinfo{year}{2013}{\natexlab{a}}).

\bibitem[{\citenamefont{Piris}(2014)}]{Piris2014c}
\bibinfo{author}{\bibfnamefont{M.}~\bibnamefont{Piris}}, \bibinfo{journal}{J.
  Chem. Phys.} \textbf{\bibinfo{volume}{141}}, \bibinfo{pages}{044107}
  (\bibinfo{year}{2014}).

\bibitem[{\citenamefont{Piris}(2017)}]{Piris2017}
\bibinfo{author}{\bibfnamefont{M.}~\bibnamefont{Piris}},
  \bibinfo{journal}{Phys. Rev. Lett.} \textbf{\bibinfo{volume}{119}},
  \bibinfo{pages}{063002} (\bibinfo{year}{2017}).

\bibitem[{\citenamefont{Mitxelena and
  Piris}(2020{\natexlab{a}})}]{Mitxelena2020}
\bibinfo{author}{\bibfnamefont{I.}~\bibnamefont{Mitxelena}} \bibnamefont{and}
  \bibinfo{author}{\bibfnamefont{M.}~\bibnamefont{Piris}}, \bibinfo{journal}{J.
  Phys. Condens. Matter} \textbf{\bibinfo{volume}{32}}, \bibinfo{pages}{17LT01}
  (\bibinfo{year}{2020}{\natexlab{a}}).

\bibitem[{\citenamefont{Mitxelena and
  Piris}(2020{\natexlab{b}})}]{Mitxelena2020b}
\bibinfo{author}{\bibfnamefont{I.}~\bibnamefont{Mitxelena}} \bibnamefont{and}
  \bibinfo{author}{\bibfnamefont{M.}~\bibnamefont{Piris}}, \bibinfo{journal}{J.
  Chem. Phys.} \textbf{\bibinfo{volume}{152}}, \bibinfo{pages}{064108}
  (\bibinfo{year}{2020}{\natexlab{b}}).

\bibitem[{\citenamefont{Piris}(2013{\natexlab{b}})}]{Piris2013c}
\bibinfo{author}{\bibfnamefont{M.}~\bibnamefont{Piris}}, \bibinfo{journal}{J.
  Chem. Phys.} \textbf{\bibinfo{volume}{139}}, \bibinfo{pages}{064111}
  (\bibinfo{year}{2013}{\natexlab{b}}).

\bibitem[{\citenamefont{Piris}(2018{\natexlab{b}})}]{Piris2018b}
\bibinfo{author}{\bibfnamefont{M.}~\bibnamefont{Piris}},
  \bibinfo{journal}{Phys. Rev. A} \textbf{\bibinfo{volume}{98}},
  \bibinfo{pages}{022504} (\bibinfo{year}{2018}{\natexlab{b}}).

\bibitem[{\citenamefont{Lopez and Piris}(2019)}]{Lopez2019}
\bibinfo{author}{\bibfnamefont{X.}~\bibnamefont{Lopez}} \bibnamefont{and}
  \bibinfo{author}{\bibfnamefont{M.}~\bibnamefont{Piris}},
  \bibinfo{journal}{Theor. Chem. Acc.} \textbf{\bibinfo{volume}{138}},
  \bibinfo{pages}{89} (\bibinfo{year}{2019}).

\bibitem[{\citenamefont{Mercero et~al.}(2021)\citenamefont{Mercero, Ugalde, and
  Piris}}]{Mercero2021}
\bibinfo{author}{\bibfnamefont{J.~M.} \bibnamefont{Mercero}},
  \bibinfo{author}{\bibfnamefont{J.~M.} \bibnamefont{Ugalde}},
  \bibnamefont{and} \bibinfo{author}{\bibfnamefont{M.}~\bibnamefont{Piris}},
  \bibinfo{journal}{Theor. Chem. Acc.} \textbf{\bibinfo{volume}{140}}
  (\bibinfo{year}{2021}).

\bibitem[{\citenamefont{Piris}(2019)}]{Piris2019}
\bibinfo{author}{\bibfnamefont{M.}~\bibnamefont{Piris}},
  \bibinfo{journal}{Phys. Rev. A} \textbf{\bibinfo{volume}{100}},
  \bibinfo{pages}{32508} (\bibinfo{year}{2019}).

\bibitem[{\citenamefont{Piris et~al.}(2013)\citenamefont{Piris, Matxain, and
  Lopez}}]{Piris2013e}
\bibinfo{author}{\bibfnamefont{M.}~\bibnamefont{Piris}},
  \bibinfo{author}{\bibfnamefont{J.~M.} \bibnamefont{Matxain}},
  \bibnamefont{and} \bibinfo{author}{\bibfnamefont{X.}~\bibnamefont{Lopez}},
  \bibinfo{journal}{J. Chem. Phys.} \textbf{\bibinfo{volume}{139}},
  \bibinfo{pages}{234109} (\bibinfo{year}{2013}).

\bibitem[{\citenamefont{Piris}(1999)}]{Piris1999}
\bibinfo{author}{\bibfnamefont{M.}~\bibnamefont{Piris}}, \bibinfo{journal}{J.
  Math. Chem.} \textbf{\bibinfo{volume}{25}}, \bibinfo{pages}{47}
  (\bibinfo{year}{1999}).

\bibitem[{\citenamefont{Piris and Mitxelena}(2021)}]{Piris2021}
\bibinfo{author}{\bibfnamefont{M.}~\bibnamefont{Piris}} \bibnamefont{and}
  \bibinfo{author}{\bibfnamefont{I.}~\bibnamefont{Mitxelena}},
  \bibinfo{journal}{Comp. Phys. Comm.} \textbf{\bibinfo{volume}{259}},
  \bibinfo{pages}{107651} (\bibinfo{year}{2021}).

\bibitem[{\citenamefont{Pritchard et~al.}(2019)\citenamefont{Pritchard,
  Altarawy, Didier, Gibson, and Windus}}]{Pritchard2019}
\bibinfo{author}{\bibfnamefont{B.~P.} \bibnamefont{Pritchard}},
  \bibinfo{author}{\bibfnamefont{D.}~\bibnamefont{Altarawy}},
  \bibinfo{author}{\bibfnamefont{B.}~\bibnamefont{Didier}},
  \bibinfo{author}{\bibfnamefont{T.~D.} \bibnamefont{Gibson}},
  \bibnamefont{and} \bibinfo{author}{\bibfnamefont{T.~L.}
  \bibnamefont{Windus}}, \bibinfo{journal}{J. Chem. Inf. Model.}
  \textbf{\bibinfo{volume}{59}}, \bibinfo{pages}{4814} (\bibinfo{year}{2019}).

\bibitem[{\citenamefont{Frisch et~al.}()\citenamefont{Frisch, Trucks, Schlegel,
  Scuseria, Robb, Cheeseman, Montgomery, Vreven, Kudin, Burant et~al.}}]{g03}
\bibinfo{author}{\bibfnamefont{M.~J.} \bibnamefont{Frisch}},
  \bibinfo{author}{\bibfnamefont{G.~W.} \bibnamefont{Trucks}},
  \bibinfo{author}{\bibfnamefont{H.~B.} \bibnamefont{Schlegel}},
  \bibinfo{author}{\bibfnamefont{G.~E.} \bibnamefont{Scuseria}},
  \bibinfo{author}{\bibfnamefont{M.~A.} \bibnamefont{Robb}},
  \bibinfo{author}{\bibfnamefont{J.~R.} \bibnamefont{Cheeseman}},
  \bibinfo{author}{\bibfnamefont{J.~A.} \bibnamefont{Montgomery},
  \bibfnamefont{Jr.}},
  \bibinfo{author}{\bibfnamefont{T.}~\bibnamefont{Vreven}},
  \bibinfo{author}{\bibfnamefont{K.~N.} \bibnamefont{Kudin}},
  \bibinfo{author}{\bibfnamefont{J.~C.} \bibnamefont{Burant}},
  \bibnamefont{et~al.}, \emph{\bibinfo{title}{Gaussian 03, \uppercase{R}evision
  \uppercase{C}.02}}, \bibinfo{note}{\uppercase{G}aussian, Inc., Wallingford,
  CT, 2004}.

\bibitem[{\citenamefont{{Johnson III}}(2020)}]{NIST}
\bibinfo{editor}{\bibfnamefont{R.~D.} \bibnamefont{{Johnson III}}}, ed.,
  \emph{\bibinfo{title}{{NIST Computational Chemistry Comparison and Benchmark
  Database, NIST Standard Reference Database Number 101, Release 21}}}
  (\bibinfo{year}{2020}).

\bibitem[{\citenamefont{{Chase, Jr.}}(1998)}]{Chase1998}
\bibinfo{author}{\bibfnamefont{M.~W.} \bibnamefont{{Chase, Jr.}}},
  \bibinfo{journal}{J. Phys. Chem. Ref. Data Monogr.}
  \textbf{\bibinfo{volume}{9}}, \bibinfo{pages}{1} (\bibinfo{year}{1998}).

\bibitem[{\citenamefont{Kramida}(2020)}]{NIST-ASD}
\bibinfo{editor}{\bibfnamefont{A.}~\bibnamefont{Kramida}}, ed.,
  \emph{\bibinfo{title}{{NIST Atomic Spectra Database, NIST Standard Reference
  Database 78}}} (\bibinfo{year}{2020}).

\bibitem[{\citenamefont{Perdew et~al.}(1982)\citenamefont{Perdew, Parr, Levy,
  and Balduz}}]{Perdew1982}
\bibinfo{author}{\bibfnamefont{J.~P.} \bibnamefont{Perdew}},
  \bibinfo{author}{\bibfnamefont{R.~G.} \bibnamefont{Parr}},
  \bibinfo{author}{\bibfnamefont{M.}~\bibnamefont{Levy}}, \bibnamefont{and}
  \bibinfo{author}{\bibfnamefont{J.~L.} \bibnamefont{Balduz}},
  \bibinfo{journal}{Phys. Rev. Lett.} \textbf{\bibinfo{volume}{49}},
  \bibinfo{pages}{1691} (\bibinfo{year}{1982}).

\bibitem[{\citenamefont{Casanova and Head-Gordon}(2008)}]{Casanova2008}
\bibinfo{author}{\bibfnamefont{D.}~\bibnamefont{Casanova}} \bibnamefont{and}
  \bibinfo{author}{\bibfnamefont{M.}~\bibnamefont{Head-Gordon}},
  \bibinfo{journal}{J. Chem. Phys.} \textbf{\bibinfo{volume}{129}},
  \bibinfo{pages}{064104} (\bibinfo{year}{2008}).
  
\expandafter\ifx\csname natexlab\endcsname\relax\def\natexlab#1{#1}\fi
\expandafter\ifx\csname bibnamefont\endcsname\relax
  \def\bibnamefont#1{#1}\fi
\expandafter\ifx\csname bibfnamefont\endcsname\relax
  \def\bibfnamefont#1{#1}\fi
\expandafter\ifx\csname citenamefont\endcsname\relax
  \def\citenamefont#1{#1}\fi
\expandafter\ifx\csname url\endcsname\relax
  \def\url#1{\texttt{#1}}\fi
\expandafter\ifx\csname urlprefix\endcsname\relax\def\urlprefix{URL }\fi
\providecommand{\bibinfo}[2]{#2}
\providecommand{\eprint}[2][]{\url{#2}}

\bibitem[{\citenamefont{Balabanov and Peterson}(2005)}]{Balabanov2005}
\bibinfo{author}{\bibfnamefont{N.~B.} \bibnamefont{Balabanov}}
  \bibnamefont{and} \bibinfo{author}{\bibfnamefont{K.~A.}
  \bibnamefont{Peterson}}, \bibinfo{journal}{J. Chem. Phys.}
  \textbf{\bibinfo{volume}{123}}, \bibinfo{pages}{064107}
  (\bibinfo{year}{2005}).

\bibitem[{\citenamefont{Quintero-Monsebaiz
  et~al.}(2021)\citenamefont{Quintero-Monsebaiz, Perea-Ram{\'{i}}rez, Piris,
  and Vela}}]{Quintero-Monsebaiz2021}
\bibinfo{author}{\bibfnamefont{R.}~\bibnamefont{Quintero-Monsebaiz}},
  \bibinfo{author}{\bibfnamefont{L.~I.} \bibnamefont{Perea-Ram{\'{i}}rez}},
  \bibinfo{author}{\bibfnamefont{M.}~\bibnamefont{Piris}}, \bibnamefont{and}
  \bibinfo{author}{\bibfnamefont{A.}~\bibnamefont{Vela}},
  \bibinfo{journal}{Phys. Chem. Chem. Phys.} \textbf{\bibinfo{volume}{19}},
  \bibinfo{pages}{2953} (\bibinfo{year}{2021}).

\end{thebibliography}
\end{document}